\documentclass[useAMS,usenatbib]{mn2e}
\usepackage{amsmath}
\usepackage{graphicx}
\usepackage{subfig}

\title[Planetesimal Formation In Self-Gravitating Discs]{Planetesimal Formation In Self-Gravitating Discs}

\author[Gibbons et al.]{P. G. Gibbons$^{1}$\thanks{E-mail: pgg@roe.ac.uk}, W. K. M. Rice$^{1}$ and G. R. Mamatsashvili$^{2,3}$\\
$^{1}$SUPA, Institute for Astronomy, Royal Observatory, Blackford Hill, Edinburgh EH9 3HJ\\
$^{2}$INAF, Osservatorio Astronomico di Torino, via Osservatorio 20,
Pino Torinese 10025, Italy\\
$^{3}$Faculty of Exact and Natural Sciences, Tbilisi State
University, Il. Chavchavadze ave. 1, Tbilisi 0128, Georgia}

\begin{document}

\date{Accepted 2011 ????. Received ????? ; in original form ?????}

\pagerange{\pageref{firstpage}--\pageref{lastpage}} \pubyear{2002}

\maketitle

\label{firstpage}

\begin{abstract}
We study particle dynamics in local two-dimensional simulations of self-gravitating accretion discs with a simple cooling law. It is well known that the structure which arises in the gaseous component of the disc due to a gravitational instability can have a significant effect on the evolution of dust particles. Previous results using global simulations indicate that spiral density waves are highly efficient at collecting dust particles, creating significant local over-densities which may be able to undergo gravitational collapse. We expand on these findings, using a range of cooling times to mimic the conditions at a large range of radii within the disc. Here we use the {\scriptsize{PENCIL CODE}} to solve the 2D local shearing sheet equations for gas on a fixed grid together with the equations of motion for solids coupled to the gas solely through aerodynamic drag force. We find that spiral density waves can create significant enhancements in the surface density of solids, equivalent to $1-10$cm sized particles in a disc following the profiles of \citet{Clarke2009} around a $\sim 1M_\odot$ star, causing it to reach concentrations several orders of magnitude larger than the particles mean surface density. We also study the velocity dispersion of the particles, finding that the spiral structure can result in the particle velocities becoming highly ordered, having a narrow velocity dispersion. This implies low relative velocities between particles, which in turn suggests that collisions are typically low energy, lessening the likelihood of grain destruction. Both these findings suggest that the density waves that arise due to gravitational instabilities in the early stages of star formation provide excellent sites for the formation of large, planetesimal-sized objects.

\end{abstract}


\section{Introduction}

There are currently two models for the formation of gas giant planets in circumstellar discs. The most widely accepted is the core accretion model \citep{Pollack1996}. Here, a core of solid material grows via a series of collisions until it becomes massive enough to accrete a gaseous envelope from the disc \citep{Pollack1996}.  This must occur before the disc is depleted of gas, a process which is observationally estimated to take $\sim10^7$ years \citep{Haisch2001}.

A major area of uncertainty in the core accretion model lies in the growth of objects from small dust grains to kilometre-sized planetesimals, which form the building blocks of planet cores. Many numerical models of the core accretion process assume that a large population of planetesimals has already formed. 

The dynamics of the smaller particles that ultimately grow to form these kilometre-sized planetesimals is, however, governed by the drag force that arises from the velocity difference between the solid particles and the surrounding gas. 

Within a few scale heights of the disc mid-plane, the pressure gradient within the disc tends to be negative. This causes the gas to orbit with sub-Keplerian velocities. The dust is not affected by the gas pressure gradient and orbits at Keplerian velocities. The drag force on the dust particles that arises from this velocity difference results in the solids losing angular momentum to the disc and drifting inward at a rate that depends on the particles' size \citep{Weid1977,Nakagawa1986}. For very small grain sizes, the dust is tightly coupled to the gas in the disc and the radial drift velocities are small. For very large objects, the solids are decoupled from the gas, move in approximately Keplerian orbits and again have very small drift velocities. Particles in the intermediate size range can, however, have large drift velocities. Although the exact size range depends on the local properties of the disc, drift velocities can exceed $10^3$cm/s for objects with sizes between 1cm and 1m \citep{Weid1977}. In order to prevent dust particles from spiralling inward into the central star before it decouples from the gas, solid material must rapidly grow from small centimetre-sized grains to large decametre sized objects. \citet{Laibe2012} do however suggest that there may be surface density and temperature profiles for which particles may survive this inward migration.

The alternative disc instability model requires that planets be able to form via direct gravitational collapse of the gaseous part of the disc, and does not require the presence of a solid core \citep{Boss1998}. Gravitational instabilities set in, for a razor--thin disc, when the sound speed $c_s$, Keplerian rotation frequency, $\Omega$ and the surface density of the disc satisfy 
\citep{Toomre1964}
\begin{equation}
Q \equiv \frac{c_s\Omega}{\pi G\Sigma}  < 1.
\end{equation}
This requires that the disc be relatively massive compared to its parent star. In the event that a disc is susceptible to such instabilities, one of two outcomes may occur. If the cooling time is long, the disc will settle into a quasi-steady state, where the cooling balances the heating generated by gravitoturbulence \citep{Gammie2001}. For short cooling times the disc may fragment, forming gas giant planets. The critical cooling time below which fragmentation occurs is commonly taken to be $t_{c,crit} =3\Omega^{-1}$ \citep{Gammie2001, Rice2003} however recent studies suggest that this threshold may not be fully converged, with recent high resolution simulations suugesting that the critical cooling time, $t_{c,crit}$ may exceed $10\Omega^{-1}$ \citep{Meru2011}. It has, however, been suggested that this result is numerical \citep{Paardekooper2011, Lodato2011, Rice2012}. \citet{Paardekooper2012} do, however, suggest that there may be an intermediate range of cooling times for which fragmentation may be stochastic, observing fragmentation in some simulations with cooling times as high as $t_c = 20\Omega^{-1}$.

Although very few Class II protostars are observed to have sufficiently massive discs to be susceptible to gravitational instabilities \citep{Beckwith1991}, observations indicate that during the Class 0 and Class I phases, massive discs may be much more common around protostars \citep{Rodriguez2005, Eisner2005,Greaves2010}. However, even though a massive circumstellar disc is thought to be present at early times in the star formation process \citep{Machida2011}, it is not clear that such discs can cool quickly enough to directly form giant planets via fragmentation.

Even if the cooling times present within discs are too long to allow giant planets to form via fragmentation, it is expected that circumstellar discs are self-gravitating in their early stages \citep{Lin1987, Lin1990}. If this is the case, these instabilities take the form of non-axisymmetric spiral structures, which have been shown to be highly efficient at transporting angular momentum outward \citep{Lodato2004,Forgan2011}. 

It has been proposed that these spiral waves may be highly effective at trapping the solids in the disc.  The gas pressure gradient changes from positive to negative across these spiral wave structures, which results in sub-Keplerian velocities on one side of the wave, and super-Keplerian on the other.  The drag force then causes dust grains to drift toward the density/pressure maxima at the centre of the waves. This can eliminate two problems currently challenging theories of planet formation. Firstly, by trapping large amounts of solid material within a spiral arm the local density of solids is much higher than the average within the disc, leading to the faster creation of planetesimals. Trapping particles in these local pressure maxima can also shield growing objects from the rapid inward drift described previously that can potentially stop large objects forming before they drift into the central object. \citet{Rice2004} showed, using global simulations, that the surface density of certain particle sizes can be enhanced by a factor of over 100 in the spiral wave structure. \citet{Rice2006} estimate that such an augmentation in the surface density of solids will lead to the creation of km-scale planetesimals through the self-gravity of the solids. \citet{Clarke2009} however noted that the simulations carried out in  \citet{Rice2004} and \citet{Rice2006} assumed a constant cooling time throughout the disc, and that the conditions required to achieve such high solid concentrations may be restricted to very large orbital radii in discs that incorporate a more realistic cooling.

Similar concentrations of particles are also seen in the presence of other density enhancements in the gas, such as those caused by magneto-rotational turbulence \citep{Johansen2006}, as well as vortices \citep{Godon2000,Klahr2003}, gaps in the gas that form due to the presence of a massive planet are also observed to create density enhancements in the solid component of the disc at their edges \citep{Paardekooper2006, Fouchet2007, Ayliffe2012}. Increased particle concentrations are thought to aid planetesimal formation in two ways, the enhanced collision rate displayed by \citet{Rice2004} will aid planetesimal formation (provided collisions are constructive). Secondly, the presence of high dust densities may actually lead to the formation of large planetesimals through direct gravitational collapse of the solids \citep{Rice2006, Johansen2007, Johansen2011}.

The ultimate goal of the present paper is an improved understanding of grain growth and planetesimal formation at early evolutionary stages when the influence of the disc's self-gravity is non-negligable. Specifically, motivated by the work of \citet{ClarkeLodato2009}, we study dynamical behaviour of particles embedded in a self-gravitating disc using a local shearing-sheet approximation. As mentioned above, \citet{ClarkeLodato2009} concluded, based on analytical estimations, that in realistic self-gravitating discs, particle agglomeration will be restricted to the outer regions of the disc (at radii of several tens of AU). At such large separations, the cooling time-scale is relatively short and hence density enhancements in spiral features are strong enough to concentrate particles on a local dynamical (orbital) time scale -- the lifetime of such features. Here, we go beyond these estimates and study in greater detail, via numerical simulations, the trapping capability of spiral waves for a wide range of disc cooling times and particle stopping times. This allows us to see how the solids will respond to the presence of spiral density waves at various radii within the disc and also understand what minimum values of gas density enhancements are required for appreciable particle concentration to occur is spiral features.

This paper is arranged as follows.  In Section 2 we outline the model we use in our simulations. In Section 3 we discuss the evolution of the gas and dust particles. Finally, we draw our conclusions in Section 4.
\\

\section[]{Model}
\label{Model}

To investigate the dynamics of solid particles embedded in self-gravitating proto-planetary discs, we solve the 2D local shearing sheet equations for gas on a fixed grid, including disc self-gravity as in \citet{Gammie2001}, together with the equations of motion of solids coupled to the gas solely through aerodynamic drag force. As a main numerical tool, we employ the {\scriptsize{PENCIL CODE}}\footnote{See http://code.google.com/p/pencil-code/}. The {\scriptsize{PENCIL CODE}} is a sixth order spatial and third order temporal finite difference code (see \citet{Brandenburg2003} for full details). The {\scriptsize{PENCIL CODE}} treats solids as numerical super-particles \citep{Johansen2006}.

To use the local shearing sheet model, we must make two simplifying assumptions, first that the disc is cool, and therefore thin $(H/r \simeq c_s/(\Omega r)\ll0.1)$, and that the disc is razor-thin, as in \citet{Gammie2001}. In the shearing sheet approximation, disc dynamics is modelled in the local Cartesian coordinate frame centred at some arbitrary radius, $r_0$, from the central object and  rotating with the disc's angular frequency, $\Omega$, at this radius. In this local frame, the $x$-axis points radially away from the central object, the $y$-axis points in the azimuthal direction of the disc's differential rotation, which in turn manifests itself as an azimuthal parallel flow characterised by a linear shear, $q$, of background velocity, ${\bf u}_0=(0,-q\Omega x)$. Our simulation domain spans the region $-L_x/2 \leq x \leq L_x/2$, $-L_y/2 \leq y \leq L_y/2$. We adopt the standard shearing-sheet boundary conditions \citep{Hawley1995}, namely for any variable $f$ we have
\begin{equation}
f(x,y,t) = f(x,y+L,t)
\end{equation}
\begin{equation}
f(x,y,t) = f(x+L,y-q\Omega Lt,t).
\end{equation}
These boundary conditions apply to all grid variables except the azimuthal component of the velocity, $u_y$, which must be adjusted to account for the relative shear between neighbouring boxes. The azimuthal velocity on the radial boundary is,
\begin{equation}
v_y(x,y,t) = v_y(x+L,y-q\Omega Lt,t)+q\Omega L.
\end{equation}
The shear parameter $q=1.5$ for Keplerian rotation profile adopted in this paper.

\subsection{Gas Density}

In this local model, the continuity equation for the gas surface density $\Sigma$ is 

\begin{equation}
\frac{\partial\Sigma}{\partial t} + {\bf\nabla}.(\Sigma{\bf u}) -q\Omega x\frac{\partial\Sigma}{\partial y}-f_D(\Sigma) = 0
\end{equation}

where ${\bf u} (u_x,u_y)$ is the gas velocity relative to the background Keplerian flow, which in our domain manifests itself as an azimuthal flow with linear shear of velocity ${\bf u}_0=(0,-q\Omega {\bf x})$. Due to the high-order numerical scheme of the {\scriptsize{PENCIL CODE}} we also include a diffusion term, $f_D$, to ensure numerical stability and capture the effects of shocks, 
\begin{equation}
f_D = \zeta_D(\nabla^2 \Sigma +\nabla \textrm{ ln } \zeta_D \cdot \nabla\Sigma).
\end{equation}
Here the quantity $\zeta_D$ is defined as,
\begin{equation}
\zeta_D = D_{sh} \langle \max_3[(-\nabla. {\bf u})_+] \rangle\Delta x^2
\label{shock}
\end{equation}
where $D_{sh}$ is a constant defining the strength of shock diffusion as outlined in Appendix B of \citet{Lyra2008}. 

\subsection{Gas Velocity}

The equation of motion for the gas velocity ${\bf u}$ relative to the background Keplerian flow ${\bf u}_0$ takes the form 
\begin{align}
\frac{\partial {\bf u}}{\partial t} + ({\bf u}\cdot\nabla){\bf u} -
q\Omega x \frac{\partial {\bf u}}{\partial y} =
& -\frac{\nabla P}{\Sigma} - 2\Omega {\bf\hat{z}}\times{\bf u}+q\Omega u_x {\bf\hat{y}} \nonumber \\
& + 2\Omega\Delta v{\bf\hat{x}} - \nabla\psi +\bf f_\nu (u),
\label{gvel1}
\end{align}
where $P$ is the two-dimensional pressure and $\psi$ is the gravitational potential of the sheet due to the gas surface density perturbation $\Sigma-\Sigma_0$. The left hand side of equation \ref{gvel1} includes terms from the velocity field {\bf u} and the Keplerian flow. The first term on the right hand side of the equation is the force due to the pressure gradient. The second and third terms represent the Coriolis effect/shear induced by the choice of coordinate system. The fourth term mimics a global radial pressure gradient, which reduces the orbital speed of the gas by $\Delta v$ and is responsible for the inward migration/drift of solids in an unperturbed disc. The main purpose for the inclusion of this term here is to see how the radial drift affects the concentration of particles by spiral density waves. The fifth term represents the effect of the gravitational potential of the disc. Finally we include an explicit viscosity term
\begin{align}
{\bf f_\nu} =&  \nu(\nabla^2{\bf u} + \frac{1}{3}\nabla\nabla\cdot{\bf u} + 2{\bf S}\cdot \nabla \textrm{ln }\Sigma) \nonumber \\
& + \zeta_\nu[\nabla(\nabla.{\bf u})+ (\nabla \textrm{ln }\Sigma +  \nabla\textrm{ln }\zeta_\nu)\nabla.{\bf u}],
\end{align}
which contains both Navier-Stokes viscosity and a bulk viscosity for resolving shocks. Here {\bf S} is the traceless rate-of-strain tensor
\begin{equation}
S_{i j} = \frac{1}{2}\left(\frac{\partial u_i}{\partial x_j}+\frac{\partial u_j}{\partial x_i} - \frac{2}{3}\delta_{i j}\nabla.{\bf u}\right)
\end{equation}
and $\zeta_{\nu}$ is the shock viscosity coefficient analogous to the shock diffusion coefficient defined in equation \ref{shock} with $\nu_{sh} = D_{sh}$ 

\subsection{Self-gravity}
The gravitational potential of the gas is found by inverting the Poisson equation for a 2D (razor thin) disc
\begin{equation}
\nabla^2\psi = 4\pi G(\Sigma - \Sigma_0)\delta(z)
\label{PoissonEqn}
\end{equation}
using the Fast Fourier Transform method outlined in the supplementary material of \citet{Johansen2007}. Here, the surface density is Fourier transformed from the $(x,y)$ plane to the $(k_x,k_y)$ plane without the intermediate co-ordinate transformation performed by \citet{Gammie2001}.  In this case a standard FFT is adapted to allow for the fact that the radial wavenumber $k_x$ of
each spatial fourier harmonic depends on time as $k_x(t) = k_x(0) + q\Omega k_yt$ in order to satisfy the shearing sheet boundary conditions \citep[see also][]{Mamatsashvili2009}. Note that since
the gravitational potential is associated with density fluctuations, only the perturbed surface density $\Sigma-\Sigma_0$ enters equation \ref{PoissonEqn}.

\subsection{Entropy}
The {\scriptsize{PENCIL CODE}} uses entropy, $s$, as its main thermodynamic variable, rather than internal energy, $U$, as used by \citet{Gammie2001}. The equation for entropy evolution is
\begin{equation}
\frac{\partial s}{\partial t} - q\Omega x\frac{\partial s}{\partial y}  + ({\bf u}.\nabla)s= \frac{1}{\Sigma T}\left(2\Sigma\nu{\bf S}^2 - \frac{\Sigma c_s^2}{\gamma(\gamma-1)t_c} + f_{\chi}(s)\right)
\end{equation}
where the first term on the right hand side is the viscous heating term and the second term is an explicit cooling term. Here we assume the cooling time $t_c$ to be constant throughout the sheet and take its value to be sufficiently large that the disc does not fragment and achieves a quasi-steady state. The final term on the right hand side is a shock dissipation term analogous to that outlined for density.

\subsection{Dust Particles}
The dust particles are treated as a number of massless numerical test-particles, using the implementation for dust particles of \citet{Johansen2007}. In this implementation dust particles have positions ${\bf x}=(x_p,y_p)$ on the grid and velocities ${\bf v}=({\rm v}_{x},{\rm v}_{y})$ relative to the unperturbed Keplerian rotation velocity ${\bf u_0}=(0,-q\Omega x_p)$ in the local frame. These are evolved as
\begin{equation}
\frac{\mathrm{d}{\bf x}}{\mathrm{d}t} = {\bf v} - q\Omega
x_p{\bf \hat{y}}
\end{equation}
\begin{equation}
\frac{\mathrm{d}{{\bf v}}}{\mathrm{d}t} = - 2\Omega
{\bf\hat{z}}\times{\bf v}+q\Omega {\rm v}_{x}
{\bf\hat{y}}+\frac{1}{\tau_f}({\bf u} - {\bf v}), 
\label{parvel}
\end{equation}
where $\tau_f$ is the friction time of the particle \citep{Johansen2006}. The first two terms in equation \ref{parvel} represent the Coriolis Force and the non-inertial force due to shear. The final term describes the drag force exerted by the gas on the particles which arises from the velocity difference between the two. Unlike the gas the particles do not feel the pressure force. The motivation for this paper is to investigate where, in a typical self-gravitating disc, the solid particles are most strongly influenced by the self-gravitating structures in the gas disc.  Additionally, we wish to investigate what range of particle sizes are most strongly affected, without making any assumption as to how particle sizes and masses are distributed. Consequently, we use massless 'test' particles. Therefore, there is no term modelling the back-reaction of the drag force in equation \ref{gvel1},  nor is the self-gravity of the particles considered in equation \ref{parvel}. We do acknowledge, however, that if the local solid density became
comparable to that of the gas, these terms would become important.

The drag force on the particles from the gas is calculated by interpolating the gas velocity field to the position of the particle, using the second order spline interpolation outlined in Appendix A of \citet{Youdin2007}. 

\subsection{Units and Initial Conditions}
We normalise our parameters by setting $c_{s,0}= \Omega=\Sigma_0= 1$. The time and velocity units are $[t] = \Omega^{-1}$ and $[v] = c_{s,0}$, resulting in the orbital period, $T = 2\pi$. The unit of length is the scale height, $[l] = H = c_{s,0}/\Omega$. We initialise the Toomre-Q parameter to be 1 throughout the sheet. This sets the gravitational constant $G$ = $\pi^{-1}$. The surface density of gas is initially set to be unity everywhere in the sheet. The box is of length $L = 80G\Sigma/\Omega^2$ and is divided into a grid of $1024^2$ cells. This choice of units sets the sheet length $L = 80H/\pi \sim 25H$. This is a much larger value for $L$ than that typically taken for simulations investigating MRI driven turbulence (e.g. \citet{Johansen2006, Johansen2011}), however is typical for those investigating instabilities due to self-gravity. \citep{Gammie2001} estimates that structure in self-gravitating discs typically arises on length scales of $\sim L/5 = 16G\Sigma/\Omega^2$. Typical proto-planetary discs are estimated to have aspect ratios $H/r = 0.05 - 0.1$ \citep{Gammie2001, Armitage2011}. This is supported by observations by \citet{Andrews2010}, who find that typical discs have scale heights ranging from $2$-$20$AU at radii of $100$AU, giving aspect ratios ranging from 0.02 to 0.2. A shearing sheet centred at $r_0 = 20$AU with a scale height of $\sim1$AU will therefore span the radial range $\sim10-30$AU. It is worth noting that the cooling time, $t_c$, which we have assumed to be constant throughout the sheet, in reality is $t_c = t_c(\Sigma,U,\Omega)$ as described by \citet{Gammie2001}. However the use of constant cooling time over a sheet of this size allows us to infer the general behaviour of the dust particles at a given location within the disc.

The gas velocity field is initially perturbed by some small random fluctuations with the rms amplitude $ \sqrt{<\delta {\bf v}^2>} = 10^{-3}$. We take the viscosity and diffusion coefficients to be $\nu = 10^{-2}$ and $\nu_{sh} = D_{sh} = 5.0$. We use $5\times10^5$ particles, split evenly between five friction times, $\tau_f = [0.01,0.1,1,10,100]\Omega^{-1}$. \citet{Bai2010} and \citet{Laibe2012} have shown there is a spatial resolution criteria which applies to coupled dust and gas simulations such as those outlined above. For the dust particles to be properly resolved, the grid spacing must satisfy $\Delta x < c_s\tau_f$. For the chosen set of parameters we have $\Delta x/c_s \sim 0.07$, they satisfying this condition for all but the $\tau_f = 0.01\Omega^{-1}$ particles. The required increase in resolution to satisfy this for the $\tau_f = 0.01\Omega^{-1}$ particles is beyond our computational limits. The effects of these particles potentially being under-resolved are discussed in section 3. Each particle species with a fixed radius is initially given zero velocity relative to the Keplerian flow, ${\bf v}(t=0)=(0,0)$, and is distributed spatially uniformly throughout the sheet.

\section{Results}
\subsection{Gas Evolution}
\label{GasEvo}

\begin{figure}
  \includegraphics[trim = 15mm 10mm 0mm 0mm, clip, width = 0.53\textwidth]{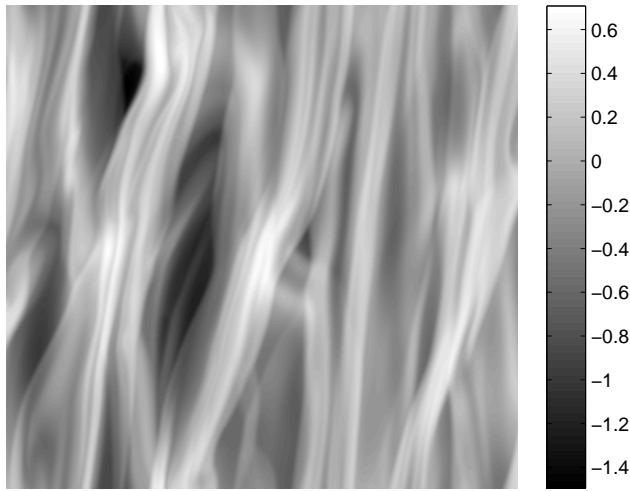}
  \caption{Logarithmic surface density of the gas after 40 orbits in the $t_c = 10\Omega^{-1}$ run. At this stage the disc is already in a quasi-steady state}
  \label{gas_dens}
\end{figure}

The evolution of the gaseous component of the disc is in good agreement with that observed in analogous studies based on the shearing sheet formalism \citep{Gammie2001,Johnson2003,Mamatsashvili2009}. The small initial velocity fluctuations grow and develop into nonlinear fluctuations in velocity, surface density and potential. Shocks then develop which proceed to heat the gas, while the cooling works to reduce the entropy of the gas. Density structures develop which are sheared out by differential rotation. These density structures tend to take on a trailing nature, which leads to a finite shear stress parameter $\alpha$ \citep{Gammie2001}. After a few orbits, the heating due to shocks is balanced by the cooling term and the thermal energy of the sheet settles into a quasi-steady self-regulated state, where the thermal, kinetic and gravitational energies of the disc are approximately constant with time. In this state, on the surface density field we clearly see elongated trailing surface density features, or density waves (Figure \ref{gas_dens}), whose particle trapping capability is our main subject of study.

The magnitude of these nonlinear density enhancements is determined by the cooling time - the smaller the cooling time, the larger the amplitude of density waves \citep{Cossins2009,Rice2011}. To analyse the efficiency of particle trapping by density wave structures at a range of radii within the disc, we consider several values of $t_c = 10, 20, 40 ,80$ and $160\Omega^{-1}$, bearing in mind that in discs realistic (radiative) effective cooling timescales tend to decrease with increasing radius \citep{Clarke2009,Rice2009}.

\subsection{Particle Concentration}

The particles, which initially are given random positions and zero velocities (relative to the Keplerian flow), are not evolved until the gas has undergone its initial burst phase and settled into a quasi-steady state. Once the gas has reached this phase, we release the dust particles. For the $t_c = 10\Omega^{-1}$ and $20\Omega^{-1}$ runs, the particles are introduced at a time $t_{par} = 10T$, while for the $t_c = 40\Omega^{-1}$, $80\Omega^{-1}$ and $160\Omega^{-1}$ runs at $t_{par} = 20T$, $t_{par} = 30T$, and $t_{par} = 50T$ respectively. In each case we evolve the system for a further 30 orbits, until the particles have also reached a quasi-steady state and come into a dynamical equilibrium with the gas - the majority of particles are trapped in density wave features, but particles aggregated within each density enhancement disperse as the the latter gets sheared out by the disc's differential rotation. These particles are then captured in a new density structure.

\begin{figure*}
  \subfloat{\includegraphics[trim = 10mm 10mm 5mm 0mm, clip, width = 0.52\textwidth]{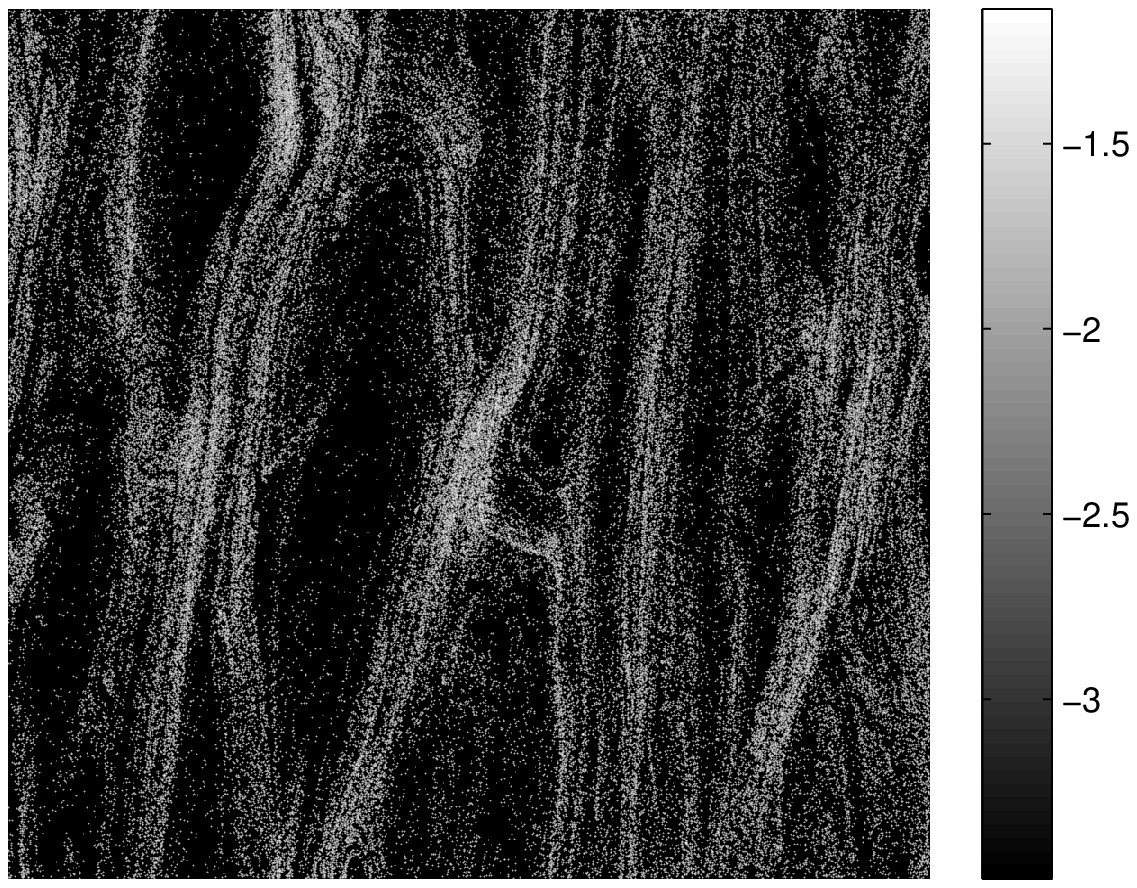}}
  \subfloat{\includegraphics[trim = 10mm 10mm 5mm 0mm, clip, width = 0.52\textwidth]{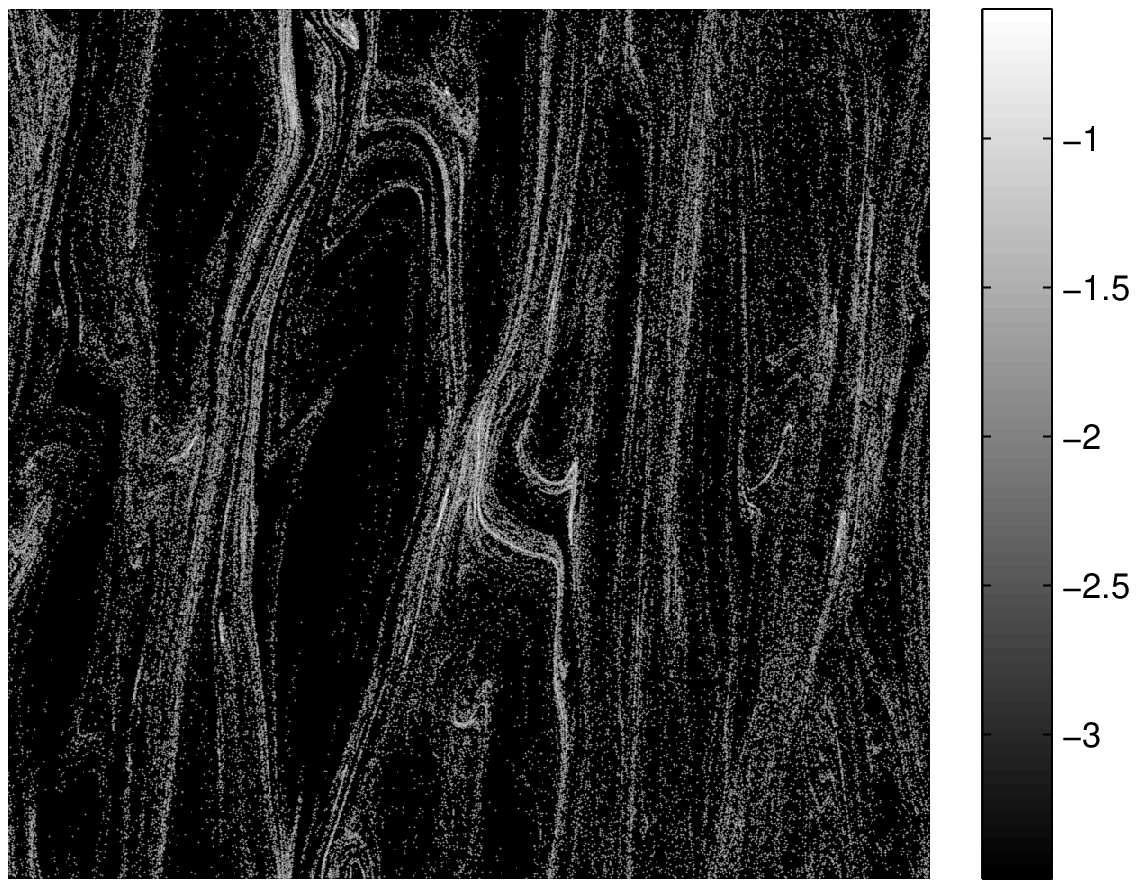}}\\
  \subfloat{\includegraphics[trim = 10mm 10mm 5mm 0mm, clip, width = 0.52\textwidth]{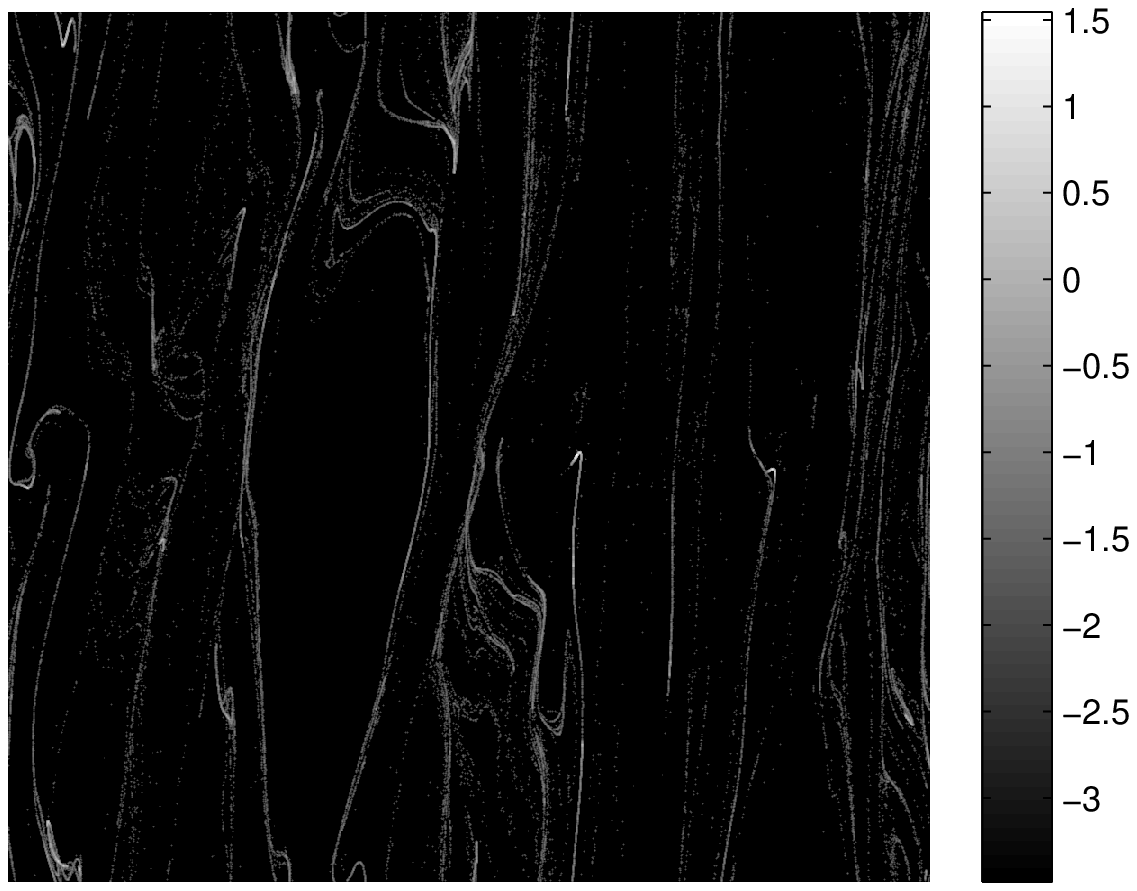}}
  \subfloat{\includegraphics[trim = 10mm 10mm 5mm 0mm, clip, width = 0.52\textwidth]{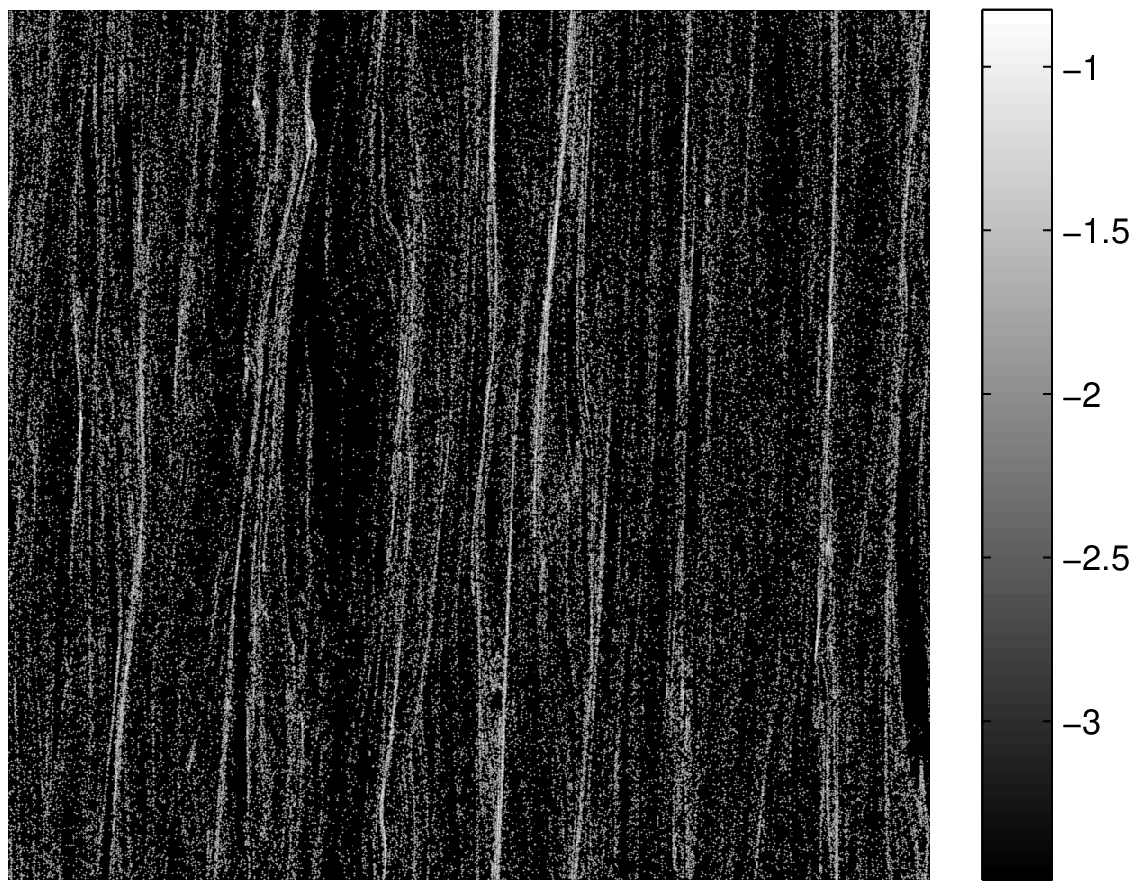}}\\
  \subfloat{\includegraphics[trim = 10mm 10mm 5mm 0mm, clip, width = 0.52\textwidth]{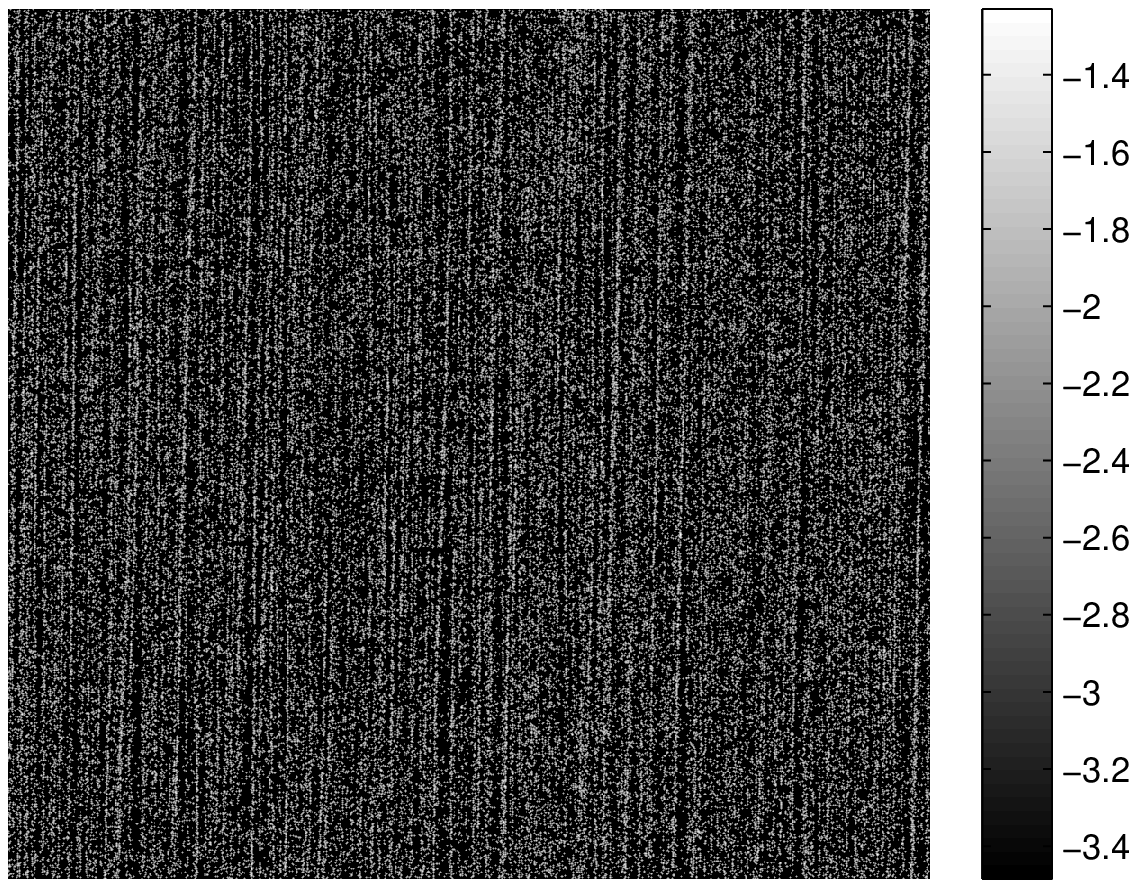}}
  \subfloat{\includegraphics[trim = 10mm 10mm 5mm 0mm, clip, width = 0.52\textwidth]{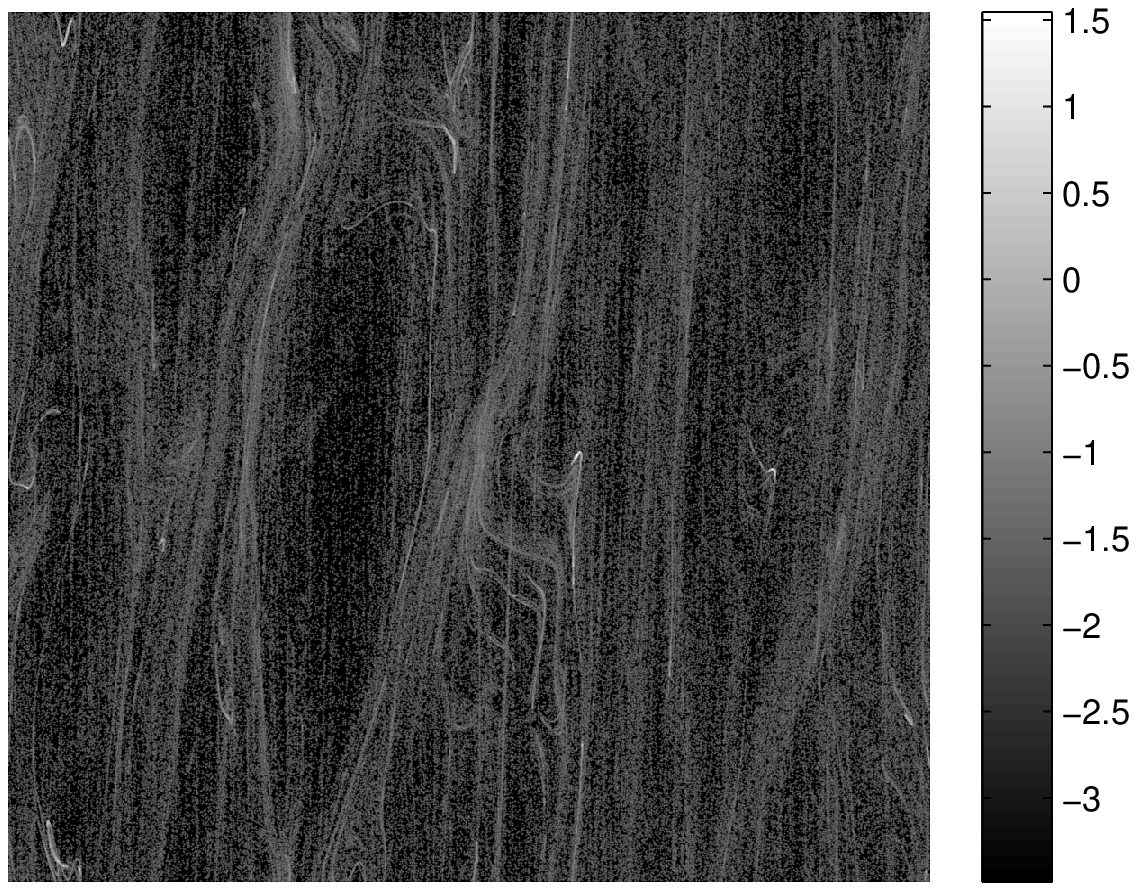}}\\
  \caption{Logarithmic surface density of the particles for the $t_c = 10\Omega^{-1}$ run at $t=40T$, when the disc is already in a quasi-steady state, 30 orbital periods after the drag force between the gas and particles has been turned on. The surface density of gas at this time is shown in Figure 1. The first five panels show the surface density of particles with friction times of $\tau_f=[0.01,0.1, 1.0, 10.0, 100.0]\Omega^{-1}$, respectively. The final panel shows the total surface density of the dust summed over all particle species. We see that intermediate-sized particles with $\tau_f \sim 1.0\Omega^{-1}$ are captured most effectively in density structures.}
  \label{surf_density}
  \end{figure*}

Upon release, the particles are drawn to local pressure maxima associated with the density waves in the gas. Figure \ref{surf_density} shows the logarithmic surface densities of the dust grains with different friction times in a quasi-steady state at $t = 40T$ in the $t_c = 10\Omega^{-1}$ run. The corresponding gas surface density at the same time is plotted in Figure \ref{gas_dens}. Comparing
Figures \ref{gas_dens} and \ref{surf_density}, we see a clear correlation between the density
enhancements in the gas and over-densities in the particles.

The degree to which the presence of spiral density waves affect the particle dynamics and the magnitude of their concentration depends on the friction time of the particles, as evident from Figures 2 and 3. The first three panels in Figure 2 show the surface densities of the smaller particles with correspondingly smaller friction times ($\tau_f = [0.01,0.1,1.0]\Omega^{-1}$ respectively). In this case, there is a high degree of correlation between the structures in the gas surface density and the arrangement of the particles, with the smallest particles (with $\tau_f \le 0.1\Omega^{-1}$, top two panels) tightly mapping the overall gas structure but the local enhancements of particle density are not exceptionally large, with typical density enhancements of no more than 10 times the mean particle density. The intermediate-sized particles (with $\tau_f =1.0\Omega^{-1}$, middle left panel) are primarily concentrated at the crests of spiral waves, tracing out the locations of the pressure maxima and reaching there the largest values of density, up to a factor of $\sim10^3$ times the mean particle density. By contrast, there is much less correlation between the surface densities of the gas and the larger particles (with $\tau_f \geqslant 1.0\Omega^{-1}$, middle right and bottom left panels) and therefore the particle over-densities are also lower. In the latter case, the mean Keplerian motion of the particles tends to dominate over the action of the drag force, reducing the effect of the latter to a small perturbation of the background motion. The specific behaviour of particles with various friction times (sizes) in the density field of spiral waves found here agrees well with that in global disc simulations of particle dynamics by \citet{Rice2004}. In this connection, we would like to mention that a similar high degree of concentration of intermediate-sized ($\sim 1m$ and $\tau_f\sim 1.0\Omega^{-1}$) dust particles also occurs in gas over-densities produced by MRI-driven turbulence in discs \citep{Johansen2006, Johansen2007}.

\begin{figure}
  \includegraphics[trim = 0mm 0mm 0mm 0mm, clip, width = 0.53\textwidth]{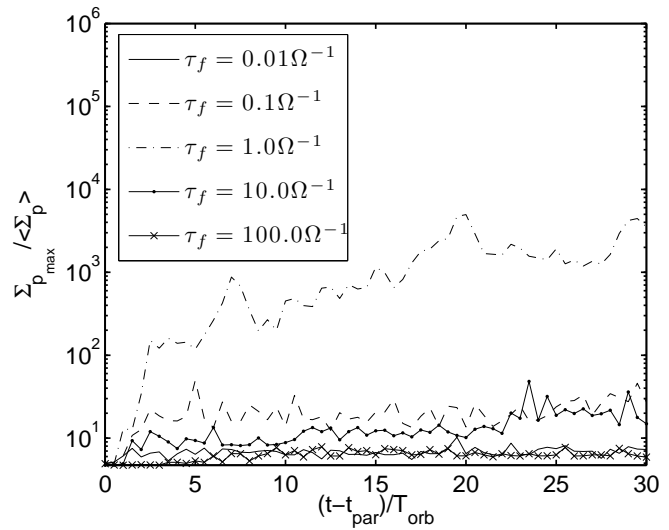}
  \caption{Maximum particle density, relative to the mean, as a function of time for a range of friction times in the $t_c = 10\Omega^{-1}$ simulation. Particles with $\tau_f=1.0\Omega^{-1}$ exhibit the highest degree of concentration in gas over-densities.}
  \label{stopping_time}
\end{figure}

Thus accumulation of the particles within spiral waves leads to an enhancement in the local surface density of the solids within the disc. Figure \ref{stopping_time} shows how the maximum value of particle density, relative to the mean particle density, inside the domain varies with time for a range of particle sizes (friction times) in the $t_c = 10\Omega^{-1}$ simulation. Particles with very short $(\tau_f \leqslant 0.1\Omega^{-1})$ and very long $(\tau_f \geqslant 10\Omega^{-1})$ friction times exhibit small density enhancements, with the maximum density of particles typically no more than a few times the mean (initial) density. However, particles that occupy an intermediate size range with $\tau_f = 0.1-10\Omega^{-1}$ exhibit orders of magnitude larger growth in gas over-densities. For example, particles with $\tau_f=0.1\Omega^{-1}$ and $\tau_f=10\Omega^{-1}$ experience an increase in concentration approximately 10 times larger than that of particles in the above mentioned first two size ranges and around 20-30 times that of the initial particle density throughout the sheet. As mentioned above, particles characterised by $\tau_f = 1.0\Omega^{-1}$ are remarkable as they exhibit the highest degree of concentration in density wave crests/maxima, with the local density of the particles typically reaching levels a factor of $\sim10^2$ times that of the mean value, or initial particle density. This range of particle density enhancements is in good agreement with analytic theory, which predicts that the effects of the aerodynamic drag force are most significant for $\tau_f = 1.0\Omega^{-1}$ particles. This consistency with analytic theory is reassuring when considering the potential non-convergence issue with the smallest $(\tau_f = 0.01\Omega^{-1})$ particles, as combined with the lack of obvious numerical problems in Figure 2, suggest that our results are correct to a good degree of accuracy, even for the insufficiently resolved particles.

In this work we have used massless `test' particles so as to focus mainly on where the self-gravitating structures in the gas disc can have the most significant affect on the
solid particles.  In a typical star forming cloud, the solid particles make up $\sim 1$ \% of the mass.  This can, however, be distributed over a wide range of particle
sizes and so it is not clear how much mass will be in particles that are strongly influenced by the gas overdensities (i.e., in particles with friction times 
$0.1\Omega^{-1} < \tau_f < 10\Omega^{-1}$).  Given that the density of these particles can be enhanced by a factor of $100 - 1000$, if $\sim 10$ \% of the solid particle mass
is in this size range, there will be regions in the disc where the gas and solid densities are comparable and we would then need to include the backreaction of the solid
particles on the gas and also the self-gravity of the solid particles to determine their subsequent evolution.  

\begin{figure}
  \includegraphics[trim = 0mm 0mm 0mm 0mm, clip, width = 0.52\textwidth]{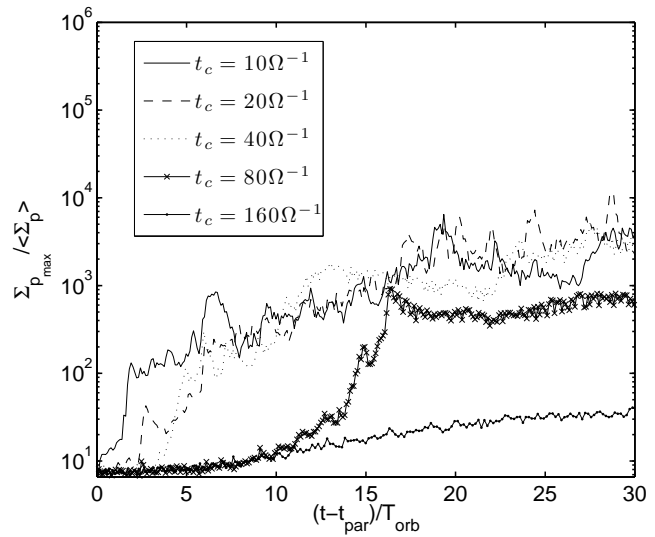}
  \caption{Maximum particle density, relative to the mean, as a function of time for a range of cooling times. Particles are somewhat more effectively concentrated in spiral density waves at short cooling times, with the $t_c = 80\Omega^{-1}$ and the  $t_c = 160\Omega^{-1}$ simulations showing a decrease in particle concentration than the shorter times considered.}
  \label{rhomax_t}
\end{figure}

The enhancement in the local density of solids is also dependent on the gas cooling time, $t_c$. As outlined in Section \ref{GasEvo}, the amplitude of spiral wave structure is determined by the value of cooling time, with smaller cooling times resulting in larger pressure gradients and higher surface densities in the spiral waves. The higher pressure gradients due to short cooling times tend to be more efficient at trapping the particles. Figure \ref{rhomax_t} shows the maximum of net surface density of all particles over time for a range of cooling times. Although in this plot we show the net particle density, at any given value of cooling time the maximum value of density is still due to the particles with $\tau_f = 1.0\Omega^{-1}$, similar to the above situation with $t_c=10\Omega^{-1}$. Here the amount of particle accumulation that occurs appears to have a dependance on $t_c$, with the longest values studied ($t_c = 80\Omega^{-1}$ and $t_c = 160\Omega^{-1}$ ) showing less concentration than that for the shorter cooling times. This decrease is due to the shallower gradients in both pressure and potential associated with the longer cooling times.  Although longer cooling times are not considered here, we expect this trend to continue, with longer cooling times exhibiting less concentration. The particles also reach a quasi-steady state much quicker in the simulations with a shorter cooling time. As the cooling time gets longer, the particles tend to form long filaments of approximately uniform density along the spiral waves in the gas, rather than tightly crowding in density wave crests.

\begin{figure}
  \includegraphics[trim = 0mm 0mm 0mm 0mm, clip, width = 0.5\textwidth]{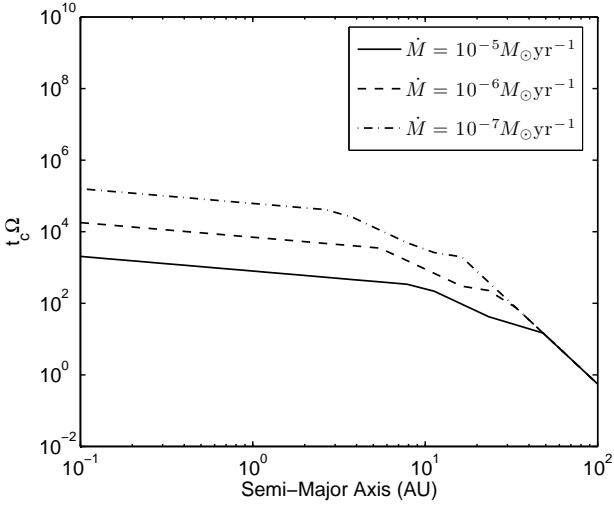}
  \caption{Analytically derived profile for $t_c$ as a function of radius for different mass accretion rates.
  The effective cooling times we consider here correspond to the radial range $\sim20-60 {\rm AU}$.}
  \label{Beta_r}
\end{figure}

It is instructive at this point to relate the parameters used in our simulations to realistic models of self-gravitating proto-planetary discs. By taking the analytically derived profile for the total stresses $\alpha$ from the appendix of \citet{Clarke2009} and finding with it an effective cooling time using the expression given in \citet{Gammie2001}, we can estimate the range of radii in the disc corresponding to the effective cooling times considered here. Figure \ref{Beta_r} shows how such obtained $t_c$ varies with radius within the disc, for given stellar accretion rates $\dot{M}$. From Fig. \ref{Beta_r} we see that the range of cooling times considered here spans a radial interval $20-60 {\rm AU}$ in the disc. This is actually the range of radii at which disc self-gravity is appreciable $(Q \sim 1)$ and maintains the disc in a quasi-steady state. At larger distances ($\geq 70 {\rm AU}$) cooling is too fast, causing the disc to fragment, whereas at smaller distances ($\leq 10 {\rm AU}$) cooling is very inefficient and so the self-gravitating density perturbations will be negligible and this region is likely to be dominated by the magneto-rotational instability \citep{Balbus1991,Zhu2009}. Once a radius within the disc is specified, we can also determine the actual values of physical variables (gas density, sound speed, etc.) in our simulations and relate the dimensionless friction times of particles used here to their real physical sizes.

The force acting on a dust particle of radius $a$ traveling at a velocity ${\bf u}$ relative to the surrounding gas of density $\rho$ is given by \citep{Whipple1972, Weid1977},
\begin{equation}
{\bf F_D} = -\frac{1}{2}C_D\pi a^2 \rho u^2 {\bf\hat{ u}},
\end{equation}
where $u=|{\bf u}|, {\bf\hat{u}}={\bf u}/u$ and the drag coefficient $C_D$ is defined as,
\begin{equation}
  C_D = \left\{
  \begin{array}{l l}
    {\displaystyle \frac{8}{3}\frac{c_s}{u}}& {\displaystyle a< \frac{9\lambda}{4}}\\
    \\
    24R_e^{-1}& R_e<1\\
    \\
    24R_e^{-0.6}& 1<R_e<800\\
    \\
    0.44& R_e>800\\
  \end{array} \right.
  \label{DragCoef}
\end{equation}
for low Mach numbers. The drag regime with $a< 9\lambda/4$ is generally called the Epstein regime, whereas the other three regimes define the Stokes drag. In expressions \ref{DragCoef}, $R_e$ is the Reynolds number characterizing gas flow in the vicinity of a dust particle, which will be defined below, and $\lambda$ is the mean free path of gas molecules. Assuming the gas to be made mainly of molecular hydrogen, for the mean free path we get
\begin{equation}
\lambda = \frac{m_{H_2}}{\rho A} = \frac{4\times10^{-9}}{\rho}{\rm cm}.
\end{equation}
The Reynolds number is given by,
\begin{equation}
R_e = \frac{2\rho a u}{\eta}
\end{equation}
where $\eta = \rho\nu$ is the gas viscosity. For collisional viscosity, we have
\begin{equation}
\eta = \frac{\rho c_s \lambda}{2}.
\end{equation}
The Reynolds number can therefore be expressed as
\begin{equation}
R_e = 4\left(\frac{a}{\lambda}\right)\left(\frac{u}{c_s}\right).
\end{equation}
Once the drag force on the particles is known, the corresponding friction time of the particles (as used in equation \ref{parvel}) can be calculated,
\begin{equation}
\tau_f = \frac{m_pu}{|{\bf F_D}|},
\end{equation}
where $m_p = \frac{4}{3}\pi a^3 \rho_p$ is the mass of a particle of internal density $\rho_p$.

By following this prescription, we estimate that for a disc with the surface density profiles derived in \citet{Clarke2009}, at the innermost radius considered ($\sim 20{\rm AU}$), the range of particle sizes considered spans 1mm-10m, with the $\tau_f =1$ particles corresponding to objects with a physical size of 10cm. At an outer radii of $\sim60$AU, the size range considered is halved, spanning 0.3mm-3m, with $\tau_f =1$ particles corresponding to 3cm size objects. We note here that the inner radii tend to be more effective at collecting `larger' particles, this allows for the possibility of small objects forming at large radii and growing as they drift inward due to aerodynamic drag.

\subsection{Particle Velocities}
\begin{figure}
  \includegraphics[trim = 8mm 12mm 10mm 7mm, clip, width = 0.49\textwidth]{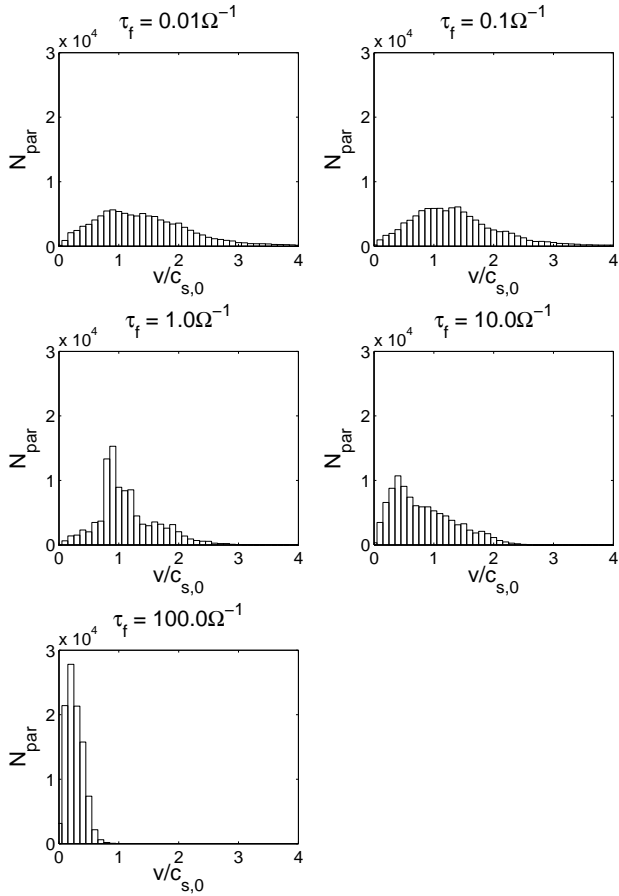}
  \caption{Histograms showing the spread of particle velocities, relative to the initial sound speed, for each friction time at the end of the $t_c = 10$ simulation. As the friction time lengthens, the distribution of velocities narrows due to the diminishing effect of the drag force. The peak of the distribution also tends to occur at lower velocities.}
  \label{taufvel} 
\end{figure}

Although large concentrations of dust and ice is vital for the growth of planetesimals, the velocities of the particles must also be considered in order to determine the evolution of the solid component of the disc.  If the relative velocities of the particles are too large, collisions become too energetic, and tend to result in large objects being broken apart, rather than growing. Figure \ref{taufvel} shows the distribution of velocities for each particle size at the end of the $t_c = 10\Omega^{-1}$ simulation. The peak of the distribution lies at higher velocities for smaller grain sizes, whilst the width of the distribution tends to also increase. In the case of the $\tau_f = 1.0\Omega^{-1}$ particles, the width of the distribution is unusually narrow. This appears to be a result of the particles being tightly concentrated at the centre of density waves, causing their velocities to become highly ordered, resulting in the majority of particles having similar velocities. This keeps the relative velocities between particles low, improving the likelihood of collisional growth occurring. 

By contrast, particles with small $\tau_f \leqslant 1.0\Omega^{-1}$ are strongly coupled to the gas and practically repeat the overall turbulent motion of gas outside spiral arms, hence the larger dispersion in the particle velocity. On the other hand, particles with large $\tau_f \geqslant 10.0\Omega^{-1}$ are very loosely coupled to the gas and therefore most of them have small velocity dispersion, i.e., they are not
effectively accelerated by gas. In Fig.6, this corresponds to the large peak in particle number near zero velocity for large friction times.

\begin{figure}
  \includegraphics[trim = 10mm 0mm 10mm 0mm, clip, width = 0.49\textwidth]{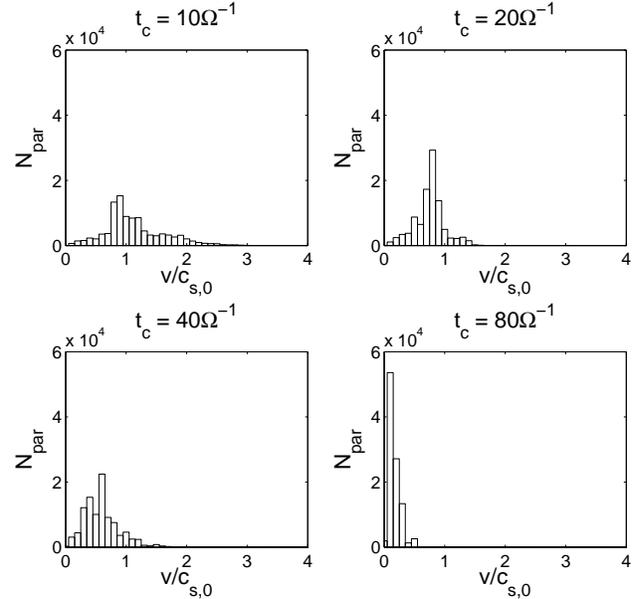}
  \caption{Histograms showing the spread of the $\tau_f = 1$ particles velocities, relative to the initial sound speed, at the end of our simulations for the four shortest cooling times considered. As the cooling time lengthens, the distribution of particle velocities narrows, with the peak of the distribution occurring at smaller velocities. 
  }
    \label{tcoolvel}
\end{figure}

As we move radially inward within the disc, we see the velocity dispersion for each particle size narrows as the cooling time increases. Figure \ref{tcoolvel} shows the velocity dispersion of the $\tau_f = 1.0\Omega^{-1}$ particles for a range of cooling times. The peak velocity decreases with increasing cooling time, and the width of the distribution narrows. We therefore see that despite the reduction in the concentration of solids that arises from the increasing cooling time, the likelihood of constructive collisions increases due to the decreasing velocity dispersion. 

As shown in figures \ref{taufvel} and \ref{tcoolvel}, the velocities of particles within our simulations are typically of order the local sound speed in the disc. For typical disc parameters, this ranges from $100-1000{\rm ms^{-1}}$. Such high velocities are typically associated with destructive collisions between particles (grain destruction is typically estimated to occur for collisions above $\sim10{\rm ms^{-1}}$), however due to the aligning of particle velocities within density waves the relative velocities between particles are typically much smaller, as demonstrated by the small spread of velocities in figures \ref{taufvel} and \ref{tcoolvel}, especially in the simulations with longer values of $t_c$. 

This opens up the possibility of a preferred region for planetesimal formation at radii between 30-40AU, where the cooling time is low enough to allow large concentrations of solids to accumulate, without the collisional velocities of particles becoming high enough to inhibit growth.

\subsection{Effects of Radial Drift}

\begin{figure}
  \includegraphics[trim = 0mm 0mm 0mm 0mm, clip, width = 0.53\textwidth]{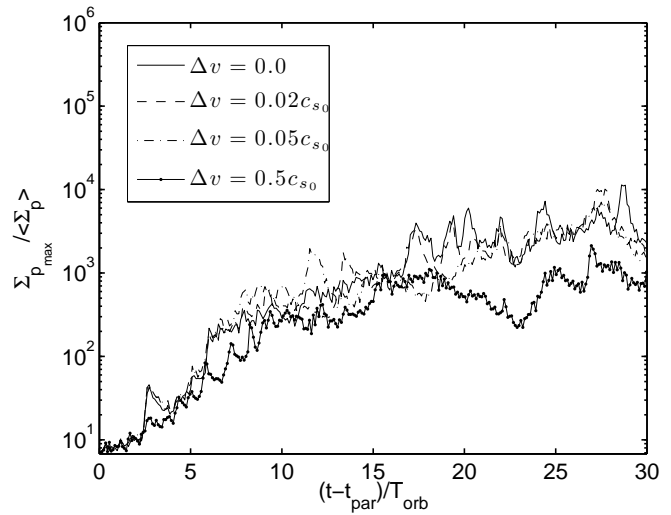}
  \caption{Maximum particle density, relative to the mean, against time for a range of drift values, $\Delta v = 0.0, 0.02, 0.05$ and $0.5$ for cooling time $t_c = 20\Omega^{-1}$. Even in the presence large amounts of radial drift, there are still significant over-densities in the particle density field due to the gravitational instabilities.} 
      \label{Sigmadrift}
\end{figure}

The above simulations were all conducted assuming that there is no radial pressure gradient supporting the disc ($\Delta v = 0$ in equation \ref{gvel1}). In reality the gas in the disc is partially supported by pressure, causing it to orbit at slightly sub-Keplerian velocities. The dust particles however travel with approximately Keplerian velocities. The resulting velocity difference results in the particles transferring angular momentum to the disc and spiralling inwards towards the star. This radial pressure gradient could potentially negate the effect of the spiral density waves, causing particles to drift through the associated pressure maxima, reducing the degree to which particles are concentrated. In order to quantify the effect of this radial drift we performed several simulations at a fixed cooling time $t_c$, varying only the value of the drift parameter $\Delta v$. Figure \ref{Sigmadrift} shows how the maximum surface density of particles evolves with time for each value of $\Delta v$ considered. In each case, we see that for typical values of $\Delta v$ (of order a few percent of the sound speed), there is very little deviation from the case with no radial drift. Even for very high drift values associated with turbulent discs, the overall concentration of particles is reduced only slightly, suggesting that the radial pressure gradient does little to change the efficiency of particle concentration within spiral waves.

\section{Discussion and Conclusions}

In this paper we present a series of simulations modelling dust dynamics in self-gravitating discs using the shearing sheet approximation. The dust particles evolve within a quasi-steady gaseous spiral structure produced by a combined effect of disc self-gravity and cooling. Particles are coupled to the gas via a drag force with a constant friction time. As in previous related studies \citep{Rice2004, Rice2006}, we find that spiral density waves in self-gravitating discs can result in significant over-densities in the solid component of the disc. A novel contribution of this work is, however, that we characterised a particle trapping capability of density waves for a wide range of disc cooling times and particle friction times. Particles with friction times $\tau_f \sim 1.0\Omega^{-1}$ tend to be affected by spiral density waves most effectively and exhibit the largest and the tightest concentrations in wave crests. At smaller friction times, the particles closely trace the overall gaseous spiral structure, however exhibit lesser concentration in wave crests. At larger friction times, particles are almost unaffected and decouple from the gas, so that they mimic the spiral structure very weakly again with smaller concentration in spiral waves. We also found that particle concentration depends on cooling time (see Fig. 4). The shorter cooling times $(t_c \leq 40\Omega^{-1})$ considered show similar amounts of particle concentration, as the spiral density waves are large enough to continually trap the majority of the particles. For longer cooling times this is no longer the case, as seen in the $t_c =80\Omega^{-1}$ and the $t_c =160\Omega^{-1}$ run, where the maximum particle density is lower. We expect this trend to continue to longer cooling times and given the rapid increase in $t_c$ with decreasing radius (Fig. 5.), do not expect disc self-gravity to play a significant role in particle dynamics at inner radii ($r<20$AU).

This implies that particle trapping by density waves is not necessarily restricted to the very outer regions of the disc, a concern raised by \citet{ClarkeLodato2009}. Generally, as mentioned above, grains with effective friction time $\tau_f \sim 1.0\Omega^{-1}$ exhibit the largest concentration in density waves. When this is translated in terms of particle size we find that grain sizes which experience the largest density enhancements depend on the radius in the disc, with larger particles more efficiently trapped at inner radii and smaller at outer radii. We also analysed the dependence of particle velocity dispersion on friction and cooling times. Given a fixed cooling time, the smaller the friction time, the larger the average velocity of particles. Conversely, at a fixed friction time, the larger the cooling time, the smaller the particle velocities are. 

Based on the trends observed in this paper, we propose the following potential route of grain growth and evolution in discs: As smaller grains can be trapped primarily at larger radii, this can lead to continual growth as grains secularly migrate inward as a result of drag forces from the gas (due to difference between Keplerian velocity of particles and sub-Keplerian velocity of gas). Small grains initially at large radii ($\sim 60 {\rm AU}$) can become trapped in gas over-densities and grow through collisions with other particles. This process can continue until the grains reach sizes larger than the optimally trapped size range ($\tau_f \sim 1.0\Omega^{-1}$). Once particles reach this size, one of two outcomes may occur. If the particles have grown sufficiently large that they are no longer strongly coupled to the gas, then they are no longer exposed to rapid inward drift, and will likely remain at that radial location for the remainder of the discs self-gravitating phase. This could result in a large number of intermediate-sized objects forming at mid-large radii within the disc. If however the particles have not reached a sufficiently large size to entirely decouple from the gas, when particles grow to a point they are no longer trapped efficiently by density waves they will start to slowly move inward in the next optimally trapped size range at smaller radii, where they will undergo further rapid growth and so on until they reach planetesimal sizes. However, this process may be partially offset at large radii, where the effective cooling time is lower, as the relative velocities between particles may be still large enough to lead to the collisional destruction of large objects. At inner radii ($\simeq 20-30 {\rm AU}$) the velocity dispersion is much lower, especially for the $\tau_f=1.0\Omega^{-1}$ particles which are clumped in spiral waves most effectively. This may lead to a ring of material, analogous to the Kuiper belt in our own Solar System, building up at intermediate radii within the disc, where the enhancement of the solids surface density is significantly augmented due to the self-gravitating structure, but the velocity dispersion is sufficiently small to avoid collisional destruction of growing objects.

Another interesting question that still needs to be addressed is the effect that large accumulations of particles have on the gas dynamics due to the back reaction of the particles on gas and the effect of particle self-gravity. As we have shown here, particle densities in the crests of spiral waves can reach 100-1000 times the mean particle density. For typical proto-planetary discs, the dust to gas mass ratio is 0.01. If there is sufficient mass in the size range that is strongly influenced by structures in the gas discs, there will be regions where the particle density can reach values comparable to, or sometimes even larger than, the local gas density. Under these conditions, it is no longer appropriate to treat the particles as massless tracer particles. To obtain a more comprehensive understanding of the particles' evolution the back reaction of particles on gas and the self-gravity of dust particles must be included. For example, in the case of planetesimal formation as a result of streaming instability, ultimate concentration of particles in the nonlinear saturated state is determined by particle feedback on gas that limits explosive growth of particle density due to linear streaming instability once the local dust-to-gas ration becomes of order unity in gas over-densities \citep{Johansen2007b}. As regards the role of particle self-gravity, we saw that once the parent gas over-density has sheared out, particles aggregated within it also dissolve, whereas particle self-gravity can in principle keep dust clumps together and prevent them from dissolving, facilitating further bonding of grains. For example, in analogous simulations of particle accumulation but in gas over-densities in MRI-driven turbulence, ultimate sticking, retainment and growth of particle aggregates occurs in fact thanks to their self-gravity \citep{Johansen2007}. We have recently run simulations that do included both the back reaction and the particles self-gravity, which suggest that the back reaction of the particles is largely negligible, and does not significantly change the evolution of the gas, even for high concentrations of dust particles where the particle density is equal to the gas, however the particle self-gravity is essential in the long term evolution of the solids, with several large, gravitationally bound structures emerging within the spiral density waves. Finally, only the 2D `shearing sheet' case has been considered in our analysis. Therefore we do not unable to observe any behaviour in the z direction, such as dust settling. Further work must be done to expand the study to the 3D `shearing box' case to study these effects.

\section*{Acknowledgments}
This work made use of the facilities of HECToR, the UKÕs national high-performance computing service, which is provided by UoE HPCx Ltd at the University of Edinburgh, Cray Inc and NAG Ltd, and funded by the Office of Science and Technology through EPSRCÕs High End Computing Programme. WKMR acknowledges support form STFC grant ST/H002380/1. The authors would like to thank Anders Johansen for help with the Pencil Code and the referee, who's comments improved the clarity of this manuscript.

\end{document}